# LOW POWER SHIFT AND ADD MULTIPLIER DESIGN

C. N.Marimuthu[1], Dr. P. Thangaraj[2], Aswathy Ramesan[3]

[1] Professor & Head / E C E, Maharaja Engineering College, Avinashi, Anna University,
*muthu_me2005@yahoo.co.in*
[2] Professor & Dean, Computer Applications, Kongu Engineering College, Perundurai, Anna University,
*ctpt@kongu.ac.in*
[3] II ME Applied Electronics, Maharaja Engineering College, Avinashi, Anna University
Tamil nadu , India
*aswathyramesan1985@gmail.com*

## ABSTRACT

*Today every circuit has to face the power consumption issue for both portable device aiming at large battery life and high end circuits avoiding cooling packages and reliability issues that are too complex. It is generally accepted that during logic synthesis power tracks well with area. This means that a larger design will generally consume more power. The multiplier is an important kernel of digital signal processors. Because of the circuit complexity, the power consumption and area are the two important design considerations of the multiplier. In this paper a low power low area architecture for the shift and add multiplier is proposed. For getting the low power low area architecture, the modifications made to the conventional architecture consist of the reduction in switching activities of the major blocks of the multiplier, which includes the reduction in switching activity of the adder and counter. This architecture avoids the shifting of the multiplier register. The simulation result for 8 bit multipliers shows that the proposed low power architecture lowers the total power consumption by 35.25% and area by 52.72 % when compared to the conventional architecture. Also the reduction in power consumption increases with the increase in bit width.*

## 1. INTRODUCTION

The power consumption in digital CMOS circuit can be described by

$$P_{avg} = P_{dynamic} + P_{shortcircuit} + P_{leakage} + P_{static} \tag{1}$$

The dynamic power dissipation is caused by charging and discharging of capacitances in the circuit. The short circuit power consumption is caused by the current flow through the direct path existing between the power supply and the ground during the transition phase. The n-MOS and p-MOS transistors used in a CMOS logic circuit commonly have non zero reverse leakage and sub threshold current. The computation of a multiplier manipulates two input data to generate many partial products for subsequent addition operations, which in the CMOS circuit design require many switching activities. The switching activities within the functional unit of a multiplier accounts for the majority of the power dissipation of a multiplier, as given in the following equation

$$P_{switching} = \alpha \, C \, V_{dd}^{2} \, f_{clk} \tag{2}$$

Where α is the switching activity parameter, C is the loading capacitance, $V_{dd}$ is the operating voltage and $f_{clk}$ is the operating frequency.

Shift-and-add multiplication is similar to the multiplication performed by paper and pencil. This method adds the multiplicand 'X' to itself 'Y' times, where 'Y' denotes the multiplier. To multiply two numbers by paper and pencil, the algorithm is to take the digits of the multiplier one at a time from right to left, multiplying the multiplicand by a single digit of the multiplier and placing the intermediate product in the appropriate positions to the left of the earlier results. To perform the entire operations for getting the final product, the conventional architecture for shift and add multipliers require many switching activities. So the dynamic power dissipation is more in conventional architecture. By eliminating or reducing the sources switching activity in the conventional multiplier, low power architecture of multiplier can be derived. Being one among the functional components of many digital systems the reduction of power dissipation in multipliers should be as much as possible. Many research efforts have been devoted for reducing the power dissipation of different multipliers (e.g.,[1]–[3], [5], [8]). Among multipliers, tree multipliers are used in high speed applications such as filters, but these require large area. The carry-select-adder (CSA)-based radix multipliers, which have lower area overhead, employ a greater number of active transistors for the multiplication operation and hence consume more power. Among other multipliers, shift-and-add multipliers have been used in many applications for their simplicity and relatively small area requirement [4].

The rest of the paper is organized as follows. Section II briefly reviews the background information about conventional shift and adds multiplier. Section III describes the architecture description of the low power multiplier. Section IV describes the low power ring counter architecture. Results are discussed in section V and conclusion is in the last section.

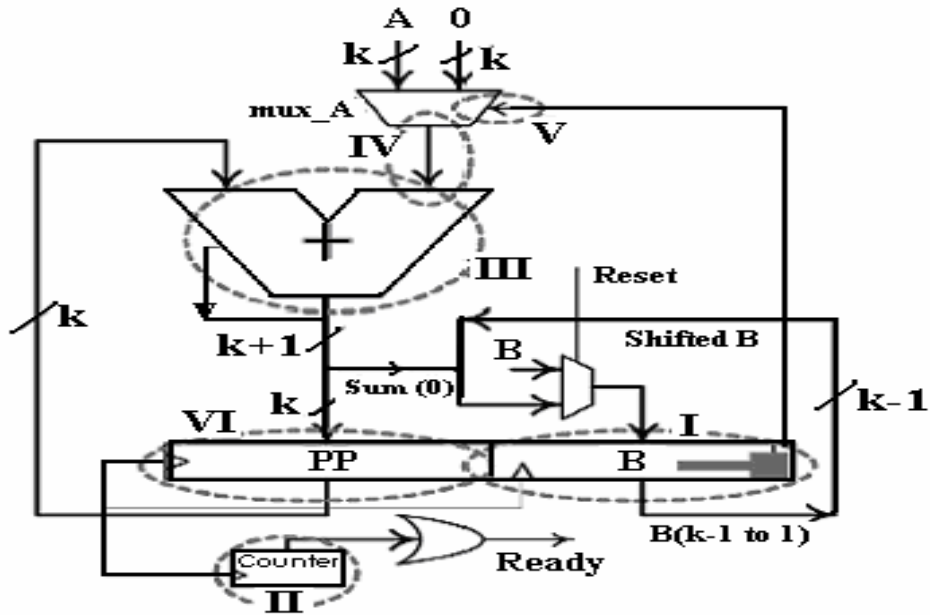

Figure 1  Architecture of conventional shift and add multiplier with major source of switching activity

## 2. SHIFT AND ADD MULTIPLIERS

Figure 1 shows the architecture of a conventional shift and add multiplier[4]. The dashed ovals show the major sources of switching activities. The multiplier is shifted in each cycle and the bit which getting out of register B is connected to the select pin of multiplexer, mux_A. As the select signal changes, the output of mux_A also changes. This causes the adder operation. The partial product is required to be shifted in every cycle. The counter is for checking whether the required number of operations has been performed. The major sources of switching activities are summarized as below

- Shifting of the 'B' register
- Activity in the counter
- Activity in the adder
- Switching between '0' and 'A' in the multiplexer
- Activity in the multiplexer select
- Shifting of the partial product register

By eliminating or reducing the switching activity described above, low power architecture can be derived.

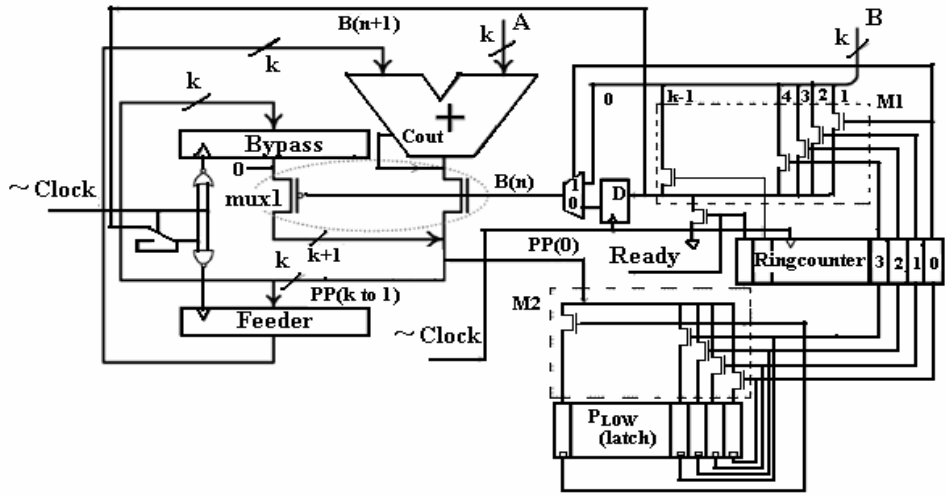

Figure 2. BZFAD architecture

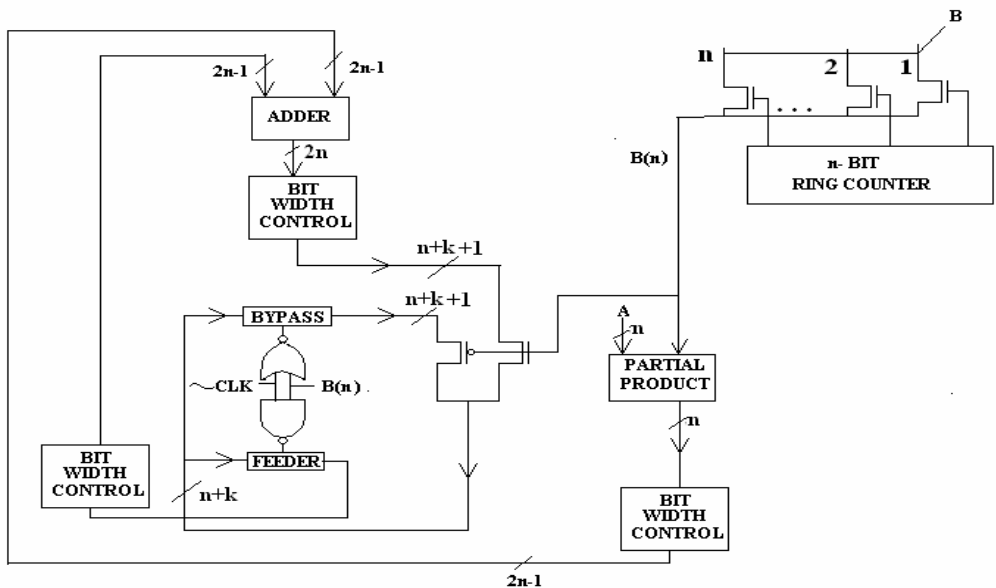

Figure 3. Proposed low power architecture

## 3. LOW POWER MULTIPLIER

The architecture in [7] BZFAD, gives an optimization in both power and area. It is shown in Figure 2. The major modifications made in [7] are , the removal of the multiplier shifting , direct feeding of multiplicand to the adder , Reduction in partial product shifting, bypassing the adder whenever possible and use of a ring counter instead of a binary counter. The proposed low power multiplier architecture is shown in Figure 3. The figure 3 is for two n bit multipliers. The value of 'k' varies between 0,1…n-1.The bit width control logic are for controlling the number of bits in each cycle.

### 3.1. Adder with registers

In conventional architecture current partial product is added to multiplicand when (0) = '1' and it is added to '0' when B (0) is '0'. Unnecessary transitions are caused by the addition of zero. Two registers called feeder and bypass are used for optimizing the operation of adder. The feeder or bypass is clocked according to the selected bit, i.e., B (n+1) in nth cycle. In each cycle the current partial product is obtained from either adder or bypass as decided by the bit B (n). Inverted clock is fed to clock gating circuit. The inverting property of the universal gates is shown in Table1.

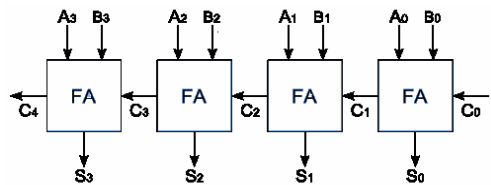

Figure 4. Ripple carry adder

The ripple carry adder is used in the proposed architecture because it has the least average transition among the look ahead, carry select, carry skip and conditional sum adders[10]. It is shown in Figure 4.

Table 1. Inverting property of universal gates

| NAND | | |
|---|---|---|
| A | B | Y |
| 1 | X | $\overline{X}$ |
| NOR | | |
| A | B | Y |
| 0 | X | $\overline{X}$ |

### 3.2. Shifting of the multiplier using ring counter

An example of a shift and add Multiplication is shown below.

```
A→      011 X
B→      010 =
        --------
        0 0 0       (B(0)=0)
         0 1 1      (B(1)=1)
        0 0 0       (B(2)=0)
        ------------
Answer→    00110
```

Figure 5. Shift and add multiplication example

In each step for generating the partial product we need to get the corresponding bit of the multiplier , in figure 5, it is shown as B(0),B(1) and B(2) . For getting the corresponding multiplier bit, the multiplier is required to be shifted. If we get the required bit without shifting, considerable power saving can be achieved. This is achieved by using a ring counter as shown in figure 6.

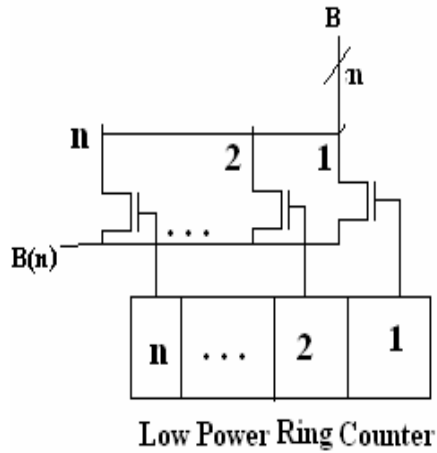

Figure 6. Multiplier with ring counter

For a 3 bit multiplier 3 bit ring counter is used. Table 2 gives the required bit and counter output combination.

TABLE 2. Counter output with required bit

| Counter output | Required bit |
|---|---|
| 001 | B(0) |
| 010 | B(1) |
| 100 | B(2) |

## 4. LOW POWER RING COUNTER

The ring counter used in the proposed multiplier is noticeably wider than the binary counter used in the conventional architecture. To minimize the switching activity of the counter we use low power ring counter.

### 4.1. Unnecessary transitions in the conventional ring counter

An n-bit synchronous ring counter is built by cascading n D flip-flops in a chain as shown in figure 7.

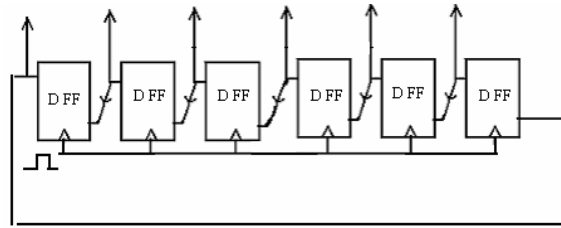

Figure 7.Conventional synchronous ring counter

In the above architecture all flip-flops have a common clock signal and each clock pulse is applied to all flip flops whereas inspecting the movement of the '1' in the counter chain reveals that each clock pulse must be applied to only 2 flip flops (not all of them).Therefore on each clock pulse   (n-2) x s unnecessary transitions are raised, in which s is the total number of transitions raised in a single flip flop and n is the number of flip flops.

### 4.2. Steps towards low power ring counter

According to the previous discussion some flip-flops can be clock gated leading to fewer switching activities. A flip-flop in a ring counter must be clocked if and only if either its input or its output is "1" immediately before the triggering clock edge comes. Therefore only 2 flip flops must be clocked in each cycle.

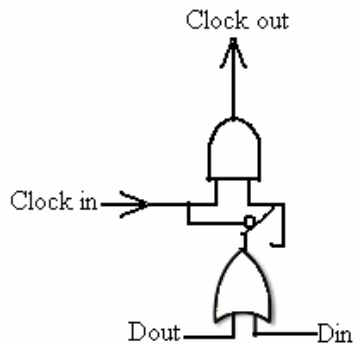

Figure 8. Clock gating logic with flip flop's input and output

The clock gating logic in the Figure 8. OR's the value of  flip flop's  input  and  output on Positive clock edges stores the result in latch. The output of  the  latch  determines  whether or not  to  gate  the  clock signal .This clock gator  is positive edge triggered. If we want to avoid all the unnecessary transitions raised by the clock signal we should provide each flip-flop with the clock gating circuitry of above figure. 8 ,but this solution ends up with a large area overhead plus due to transitions in clock gator themselves the resulting ring counter will not have fewer switching activity. A better solution is used in the low power multiplier architecture.

One of the important properties of the ring counter is that its output is one hot encoded. This property of the ring counter makes its output wide especially as the counter size increases. To reduce the switching activity of the counter the counter is partitioned in to a number of blocks which are clock gated with a special clock gating structure whose power and area overheads are independent of the block size, controlling with the low power ring counter helps to get a low power low area architecture,thus avoids the trade off between power and area([6],[9],[5]). The clock gating structure is shown below Figure10

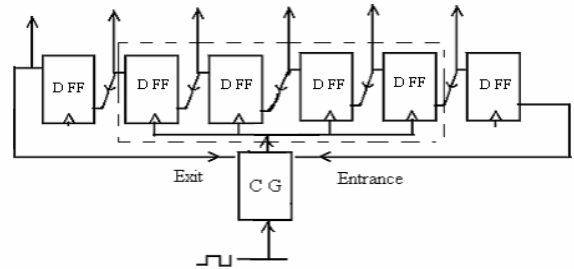

Figure 9. Low power architecture for ring counter with block of size 4

.

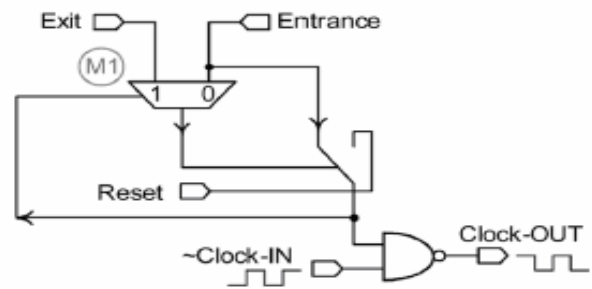

Figure 10. Clock gating structure for the low power ring counter

## 5. RESULTS AND DISCUSSION

In this paper, we propose a low-power, low-area architecture for shift and add multipliers. The low power architecture avoids the unwanted switching activities and thus minimizes the switching power dissipation. The conventional and proposed design of 8 bit multiplier can be verified using Modelsim 6.5 with VHDL code and it is shown in Figure 11 and Figure 12.The power consumption is analyzed using Xilinx software and its reports are shown in Figure 13 and Figure 14.

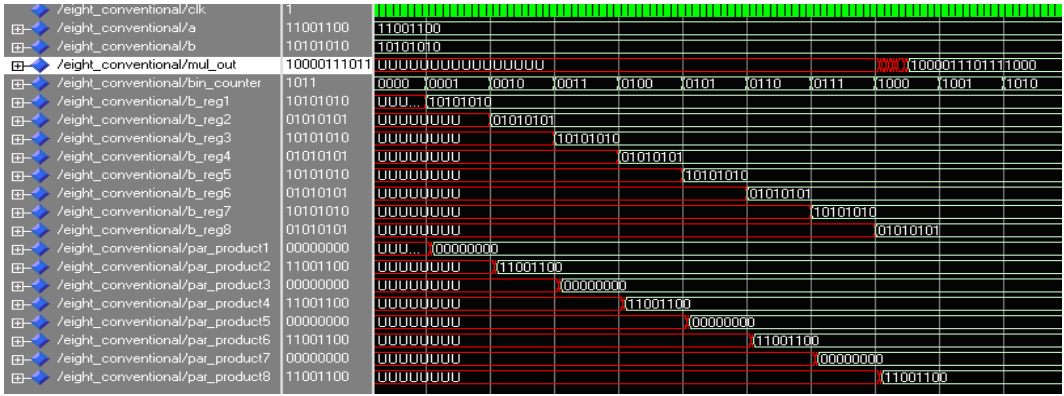

Figure 11. Simulation results for conventional 8 bit multiplier

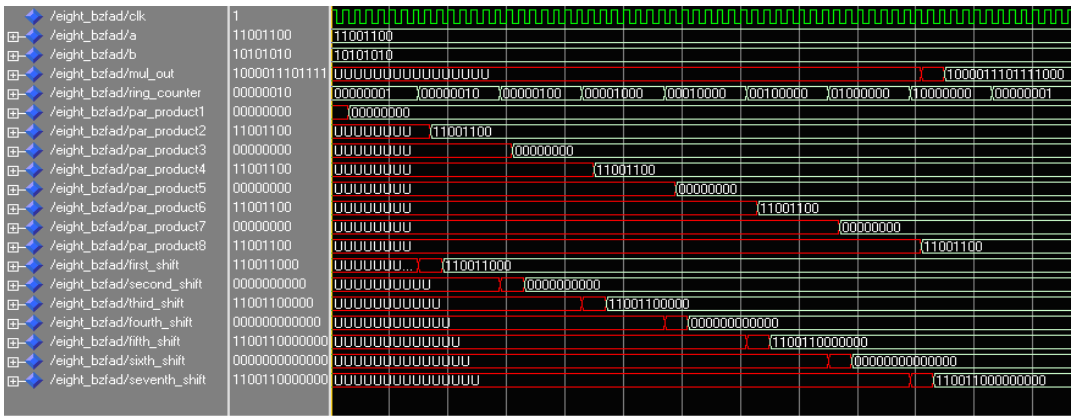

Figure 12. Simulation results for 8 bit multiplier using proposed architecture

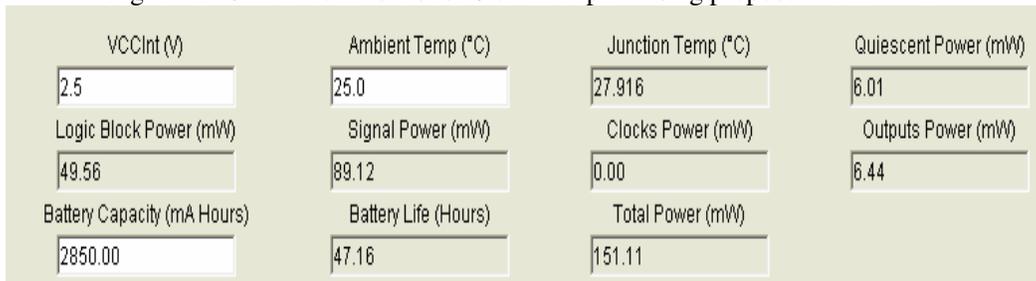

Figure 13. Power report for conventional 8 bit multiplier

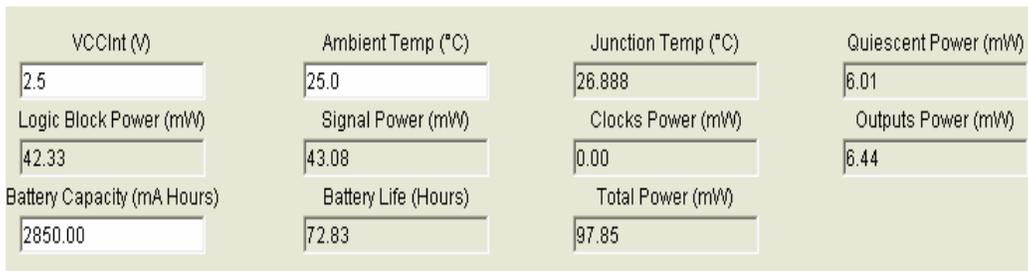

Figure 14. Power report for proposed 8 bit multiplier

| Simulated Area report for conventional 8 bit multiplier is given below |
|---|
| Design Summary |
| Number of errors:     0 |
| Number of warnings:   0 |
| Number of Slices:            662 out of  1,200   55% |
| Simulated Area report for proposed 8 bit  multiplier is given below |
| Design Summary |
| Number of errors:     0 |
| Number of warnings:   0 |
| Number of Slices:            313 out of  1,200   26% |
| Simulated Delay report for conventional 8 bit multiplier is given below |
| Timing Summary: |
| Minimum period: 52.071ns (Maximum Frequency: 19.205MHz |
| Simulated Delay report for   proposed 8 bit multiplier is given below |
| Timing Summary: |
| Minimum period: 48.812ns (Maximum Frequency: 20.487MHz) |

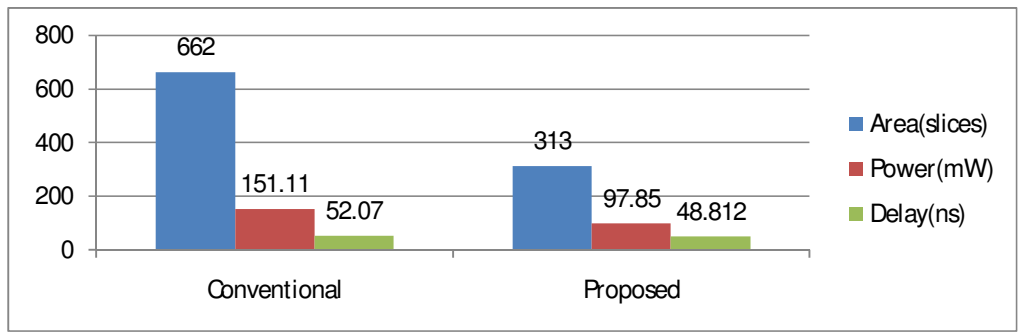

Figure 15. Power Area and Delay comparison for conventional and Proposed 8 bit multiplier

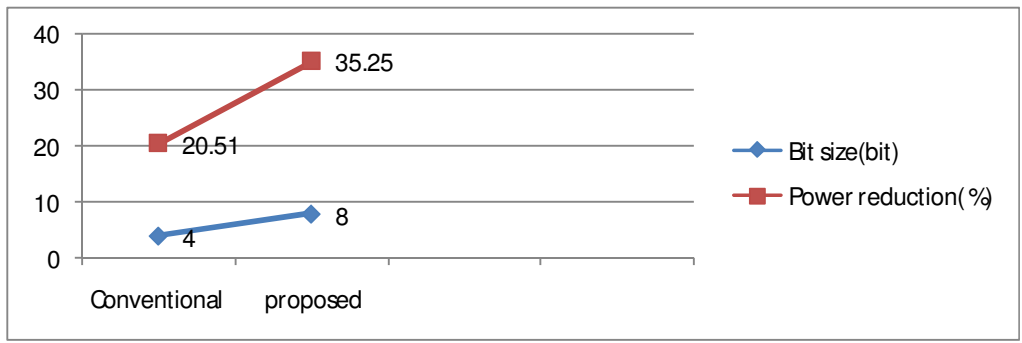

Figure 16. Relationship between Power reduction and Bit size of Multiplier

The comparison of synthesis for the two multipliers is shown in table 3

TABLE 3
Synthesis report for two multipliers

| Multiplier Type | Conventional shift and add multiplier | proposed architecture |
|---|---|---|
| Vendor | Xilinx | Xilinx |
| Device and Family | Spartan 2 | Spartan 2 |
| Estimated area | 662 slices | 313 slices |
| Power dissipation | 151.11mW | 97.85mW |

TABLE 4
Increase in power reduction with bit width

| Bit width | Reduction in power consumption |
|---|---|
| 4 | 20.51% |
| 8 | 35.25% |

## 6. CONCLUSION

The proposed architecture lowers the power dissipation and area when compared to a conventional shift and add multiplier shown in Figure 15 . A multiplexer with one hot encoded bus selector is used for avoiding the switching activity due to the shifting of the multiplier register. Feeder and bypass registers are used for avoiding the unnecessary additions. The proposed architecture makes use of bit width control logic and a low power ring counter .The design can be verified using Modelsim 6.5 with VHDL code and power consumption is analyzed using Xilinx software. From Table 3 and Figure 15 , proposed architecture can attain 35.25% power reduction and 52.75% area saving when compared to the conventional shift and add multipliers. Also from Table 4 and Figure 16 reduction of power consumption in multipliers can be increases with the increase in bit width of operands, whereas in design [5] reduction in power consumption decreases with the increase in bit width.